\documentclass[a4paper,fleqn,usenatbib]{mnras}
\usepackage[T1]{fontenc}
\usepackage{ae,aecompl}
\usepackage{graphicx}
\usepackage{amssymb}
\usepackage{amsmath}
\usepackage{hyperref}
\usepackage{tabularx}

\usepackage[usenames,dvipsnames]{color}

\newcommand{\dom}[1]{{\bf\textcolor{black}{{#1}}}}

\title[Emulating reionization simulations]{
A deep learning model to emulate simulations of cosmic reionization
}
\author[Chardin et al.]
{\parbox{\textwidth}{Jonathan Chardin,$^{1}$\thanks{E-mail: jonathan.chardin@astro.unistra.fr}
Gr\'egoire Uhlrich,$^{1,2}$
Dominique Aubert,$^{1}$ 
Nicolas Deparis,$^{1}$
Nicolas Gillet,$^{3}$
Pierre Ocvirk,$^{1}$ and
Joseph Lewis$^{1}$
}\vspace{0.4cm}
\\
$^1$Observatoire Astronomique de Strasbourg, Universit\'e de Strasbourg, CNRS UMR 7550, 11 rue de l\textquoteright Universit\'e, F-67000 Strasbourg, France \\
$^2$IPNL, Universit\'e de Lyon, Universit\'e Lyon 1, CNRS/IN2P3, 4 rue E. Fermi 69622 Villeurbanne cedex, France\\
$^3$Scuola Normale Superiore, Piazza dei Cavalieri 7, I-56126 Pisa, Italy}

\begin{document}


\date{Accepted / Received }

\pagerange{\pageref{firstpage}--\pageref{lastpage}} \pubyear{2019}

\maketitle

\begin{abstract}

We present a deep learning model trained to emulate the radiative transfer during the epoch of cosmological reionization. CRADLE (Cosmological Reionization And Deep LEarning) is an autoencoder convolutional neural network that uses two-dimensional maps of the star number density and the gas density field at z=6 as inputs and that predicts 3D maps of the times of reionization $\mathrm{t_{reion}}$ as outputs. 
These predicted single fields are sufficient to describe the global reionization history of the intergalactic medium in a given simulation.
We trained the model on a given simulation and tested the predictions on another simulation with the same paramaters but with different initial conditions. 
The model is successful at predicting $\mathrm{t_{reion}}$ maps that are in good agreement with the test simulation. 
We used the power spectrum of the $\mathrm{t_{reion}}$ field as an indicator to validate our model. We show that the network predicts large scales almost perfectly but is somewhat less accurate at smaller scales. While the current model is already well-suited to get average estimates about the reionization history, we expect it can be further improved with larger samples for the training, better data pre-processing and finer tuning of hyper-parameters.
Emulators of this kind could be systematically used to rapidly obtain the evolving HII regions associated with hydro-only simulations and could be seen as precursors of fully emulated physics solvers for future generations of simulations.

\end{abstract}



\section{Introduction}
\label{intro}

The process of cosmic reionization is the period that sees the cosmic hydrogen content of the intergalactic medium (IGM) being progressively ionized by the first sources of ionizing radiation during the first billion years of cosmic history
(\citealt{2000ApJ...535..530G}, \citealt{2001PhR...349..125B}, \citealt{2005MNRAS.361..577C} and \citealt{2018PhR...780....1D} for a recent review). 
This process marks the last major transition for cosmic gas in the history of the Universe and is of prime importance to explain what happened to the next generation of galaxies and to understand the Universe we see today at $z=0$.

Correctly modelling this phenomenom in order to interpret future observational results is 
one of the upcoming challenges in astrophysics.
With the promise of new facilities dedicated to the study of this epoch with instruments like SKA (see \citealt{2016MNRAS.461.3135B} and \citealt{2016JApA...37...27D}) or JWST (see \citealt{2006NewAR..50..113W} and \citealt{2019arXiv190306027W}), the community wants to be ready to investigate the parameter space from the theoretical side. 
This can be done with a variety of models ranging from analytical (\citealt{2000ApJ...534..507C}, \citealt{2004ApJ...613....1F}, \citealt{2006MNRAS.369.1055B} and \citealt{2009MNRAS.394..960C}), semi-numerical (\citealt{2007ApJ...654...12Z}, \citealt{2009ApJ...703L.167A}, \citealt{2009MNRAS.393...32T} and \citealt{2011MNRAS.414..727Z}), to full simulations (\citealt{2001NewA....6..437G}, \citealt{2006MNRAS.369.1625I}, \citealt{2011MNRAS.417L..93O}, \citealt{2013MNRAS.436.2188R},  \citealt{2013ApJ...777...51O}, \citealt{2014ApJ...793...29G}, \citealt{2014A&A...568A..52C}, \citealt{2015MNRAS.454.1012A}, \citealt{2016MNRAS.463.1462O}, \citealt{2018arXiv181111192O} and \citealt{2018ApJ...856L..22A}) incorporating an increasingly accurate description of the physics at play during the epoch of reionization.

Simulations of cosmic reionization are computationaly expensive because of the necessary inclusion of radiative transfer physics : propagation at the speed of light and out-of-equilibrium thermo-chemistry induce short timescales, leading to large amounts of calculations to cover the first billion years in the Universe history. Hardware acceleration, with e.g. GPUs, can reduce their cost (see e.g.  \citealt{2010ApJ...724..244A}, \citealt{2016MNRAS.463.1462O}) but requires dedicated devices, available only in limited numbers or in specific supercomputing facilities, even though their usage is becoming more widespread thanks to the rise of machine learning. 
Another way to accelerate such calculations is to use the so-called reduced speed of light approximation (see e.g. \citealt{rosdahl_ramses-rt:_2013}, or \citealt{2017MNRAS.468.4831K} for an extension of this technique to variable speed of light). With typical values of $\mathrm{\tilde{c}=[0.01-0.1]\times c}$, computing times can be divided by factors ranging from 10 to 100.
However, even with such performances, such simulations remain costly  and can introduce spurious artefacts compared to simulations using the actual speed of light (\citealt{2016ApJ...833...66G}, \citealt{2019A&A...622A.142D} and \citealt{2018arXiv180302434O}). 
Overall, even with such techniques, simulations of cosmic reionization are still challenging.

In this paper we propose to use the recent advent of deep learning methods to reassess this issue.
Machine learning algorithms are powerful in the sense that they can learn complex relationships in data, without requiring any prior functional form to describe a particular physical problem.
Due to the increasingly large amounts of data encountered in astronomy, be it through observations or simulations, the use of machine learning techniques is becoming more and more widespread.
For example, \citet{2016MNRAS.455..642K} and \citet{2016MNRAS.457.1162K} used a large set of semi-analytical and full hydrodynamical data obtained from simulations to learn the underlying physics governing the galaxy formation process. \citet{2018MNRAS.477.1484U} and \citet{2019MNRAS.483.1295U} used supervised learning to infer physical properties of galaxies from their emission-line spectra.
Among others, \citealt{2018A&A...611A...2S} used convolutional neural network  to build a strong gravitational lens detectors while \citealt{2018MNRAS.476.1151P} used the same kind of technique to characterize strong absorbers of neutral hydrogen (i.e. damped Ly$\alpha$ systems) in quasar spectra.

\citet{2019arXiv190210159N} recently reviewed what has been done in the field of cosmology with the advent of machine learning techniques. 
Among them, studies were undertaken to address the epoch of reionization in the context of deep learning. \cite{2017MNRAS.468.3869S} used such methodologies on synthetic 21cm power spectra to extract physical properties of the reionization process.
\cite{2019MNRAS.484..282G} first proposed to use light cones of the 21cm surface brightness as input of convolutional neural network  to predict cosmological parameters. 
In the same spirit, \citealt{2019MNRAS.483.2524H} used synthetic 21 cm light cones drawn from simulations to predict what  the relative contribution to reionization between star-forming galaxies and AGNs is.

With this study, we aim to go beyond the aformentioned works, to predict physical fields relevant to reionization from other physical fields.
Our aim is to use fields of gas density and star number counts  as inputs of a neural network to predict maps of reionization times $\mathrm{t_{reion}}$. The reionization times maps encode the whole reionization history of a given simulation : having a neural network predictor would allow to assign locations of HII bubbles at all times in simulations without radiative transfer, which would in turn make possible the quick acquisition of e.g. a mean reionization history associated with those simulations. 

We propose to use actual radiative-hydrodynamics simulations of cosmic reionization to feed the learning process of such networks.
To some extent, we aim at designing a tool similar to semi-analytical models, but rather than using an explicit model we propose to create an implicit model. Such a model would be provided by full-physics simulations and would constitute rather a 'semi-numerical' model, orders of magnitude faster than the simulations it originates from.
More generally, one can envision deep learning methods to emulate physics solvers (i.e. coupled differential equations solvers), using simulations' products as training models but with a much smaller execution times than for actual simulation codes : the radiative transfer case used here should merely be seen as an example of a much greater potential. 

This paper is organized as follows. We first present the simulation of cosmic reionization used in this study. Second, we detail the architecture and the training performance of our neural network in Sect. \ref{autoencoder} before giving the strengths of this model in Sect. \ref{validation}. We finally discuss how our results could be improved and generalized in the near future in Sect. \ref{discussion}.

 \begin{figure*}
   \begin{center}
      \includegraphics[width=\textwidth]{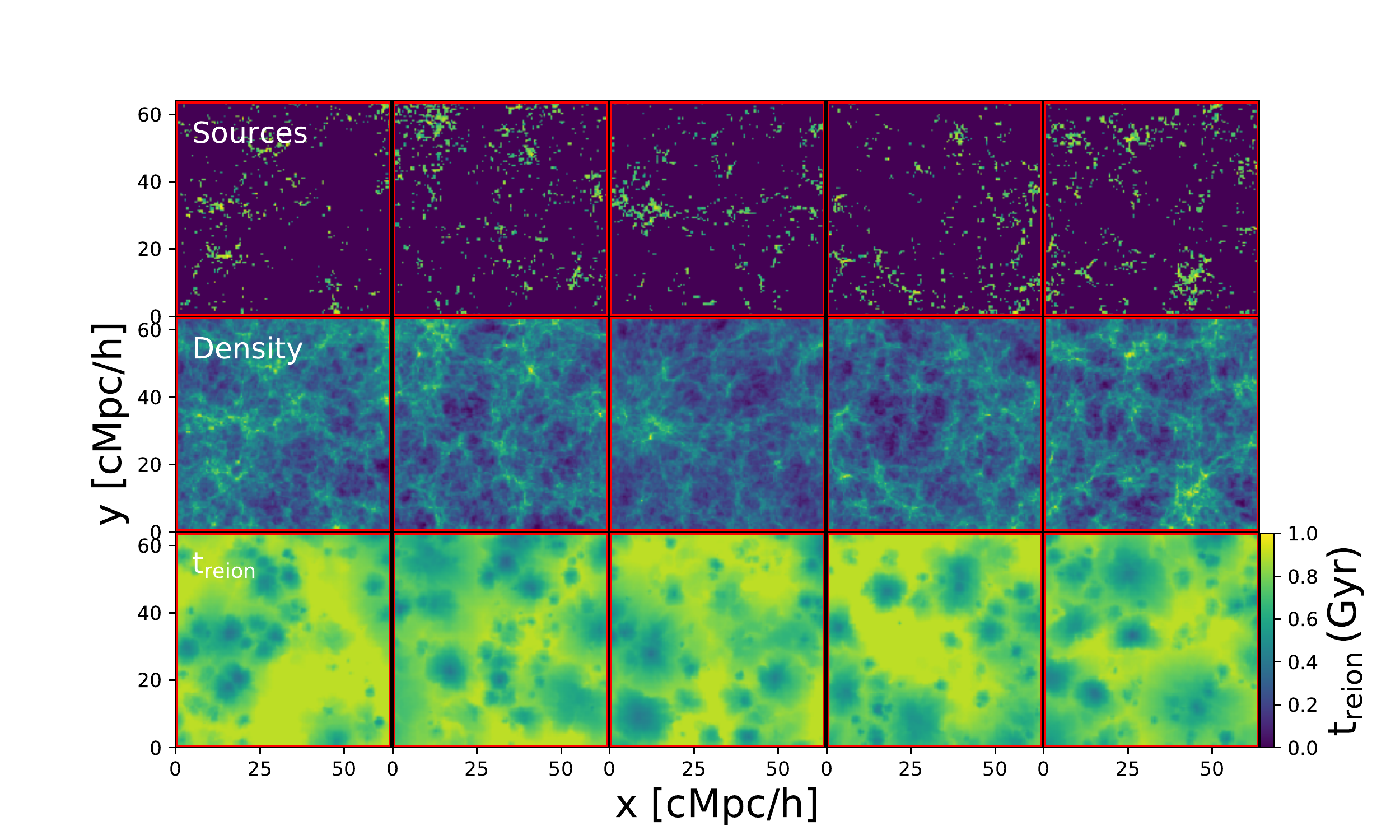}   
  \caption{Example of fields used to train our neural network. The stellar particles number counts and gas density fields are used as inputs of the network and the $\mathrm{t_{reion}}$ field is what should be predicted.
  The stellar and the gas density fields are pre-processed as described in Sect. \ref{dataset} while the $\mathrm{t_{reion}}$
  field is not touched. Briefly, they are zero-centered as well as unit variance transformed. Moreover, the stellar and gas density fields are Gaussian smoothed in the transverse direction of the plane seen on the figure to keep three-dimensional information in two dimensions.
  We can see at first glance that the three fields are correlated. In principle, the neural network should be able to infer this underlying correlation.}
    \label{ICsauto}
  \end{center}
 \end{figure*}

\section{Simulations of cosmic reionization with EMMA}
\label{EMMA_simu}

In this work, cosmological simulations of the Reionization were produced using the EMMA simulation code (\citealt{2015MNRAS.454.1012A}) : the code tracks the collisionless dynamics of dark matter, the hydrodynamics of baryons, star formation and feedback and the radiative transfer using a moment based method (see e.g. \citet{2018ApJ...856L..22A} \citealt{2019A&A...622A.142D}). This code adheres to a eulerian description, with fields described on grids, and enables adaptive mesh refinement techniques to increase the resolution in collapsing regions.

For the current study, we used an existing pair of large scale, well-resolved simulations (with high enough resolution to follow the densest absorbers that are subject to self-shielding as shown in \citealt{2018MNRAS.478.1065C}). The two simulations share the same parameters, but with different displacement phases in the initial conditions. In both cases, the (128 Mpc/h)$^3$ volume is sampled with $1024^3$ cells at the coarsest level. Refinement is triggered when the number of dark matter particles exceeds 8, up to 6 refinement levels. 

A Planck 2015 cosmology was used (\citealt{2015arXiv150201589P}) to generate the initial conditions, with a starting redshift of z=150. Simulations were stopped at z=6. 
The dark matter mass resolution is $2.1\times 10^8 M_\odot$ and the stellar mass resolution is $6.1\times10^5 M_\odot$. Star particles produce ionizing radiation for 3 Myrs, with an emissivity provided by the Starburst99 model for a Top-Heavy initial mass function and a Z=0.001 metallicity. Star formation proceeds according to standard recipes described in \citealt{RAS06}, with an overdensity threshold equals to 20 to trigger the gas-to-stellar particle conversion with a 0.1 efficiency : such values allow the first stellar particles to appear at $z\sim 17$. Supernovae feedback follows the prescription used in \citealt{2018ApJ...856L..22A} : as they reach an age of 15 million years, stellar particles dump 9.8$\times 10^{11}$ J/stellar kg in the surrounding gas, 1/3 in the form of thermal energy, 2/3 in the form of kinetic energy.

Using these parameters, we obtain a cosmic star formation history consistent, but somewhat under-estimated , with constraints by \citealt{bouwens_uv_2015}. 
Even if these simulations are not fully consistent with observations, it is more than enough for the purpose of this paper, which is to demonstrate the idea that a machine learning algorithm can learn the physics of reionization and can be used to predict the ionization field of non RT-simulations. 
Hence, the technique exposed here could be generalized in the future on simulations more carefully calibrated with observations.

This pair of simulations was produced on the Occigen supercomputer (CINES, France) on a standard CPU architecture :  EMMA GPU acceleration capabilities were not enabled and a reduced speed of light $\tilde c=0.1 c$ has been used to reduce the cost of radiative transfer. For the purpose of the current investigation, we did not use the simulation products at full resolution : outputs were degraded to a $256^3$ resolution to fit within the capabilities of our hardware dedicated to neural network training.

These two simulations will be labeled respectively as TESTSIM and TRAINSIM. TRAINSIM is the simulation used for the actual training of our model, whereas TESTSIM is used to quantify its predicting power. TESTSIM is never used during the training process and thus provides a way to test the model on a completely independent dataset.

\section{Autoencoder convolutional neural network}
\label{autoencoder}

\subsection{Outputs : $\mathrm{t_{reion}}$ fields}
\label{treion_explanation}

Our aim with this study is to predict the 3D $\mathrm{t_{reion}}$ field of a simulation, also known as 'reionization maps', built by marking cells with the cosmological time at which it crosses a given ionization fraction threshold. 
At the end of a simulation, it provides the full reionization history and this field can be used to reconstruct the HII regions' spatial distribution at all cosmic times (see \citealt{2013ApJ...777...51O}, \citealt{2018ApJ...856L..22A} and \citealt{2019A&A...622A.142D}). 
Predicting such a field with a neural network should be very useful for those who only have hydro-simulations at their disposal to  get rough estimates of the mean reionization history of their simulations.
In our case, $\mathrm{t_{reion}}$ maps are built on the fly by the EMMA simulation code. We choose an ionization fraction threshold $\mathrm{x_{HII}\ge0.5}$ to consider cells of the simulation as ionized and to mark them with the corresponding cosmological time of reionization $\mathrm{t_{reion}}$. 
Varying this threshold from $\mathrm{x_{HII}\ge0.5}$ to $\mathrm{x_{HII}\ge0.9}$ is unlikely to affect the shape and size of HII regions (see \citealt{2012A&A...548A...9C} Appendix A.3) and thus is unlikely to affect the  spatial distribution of $\mathrm{t_{reion}}$ maps.
Therefore varying this threshold should not impact what the neural network will learn in the next sections.

\subsection{Inputs and Data set preprocessing}
\label{dataset}

\begin{figure*}
   \begin{center}
      \includegraphics[width=\textwidth]{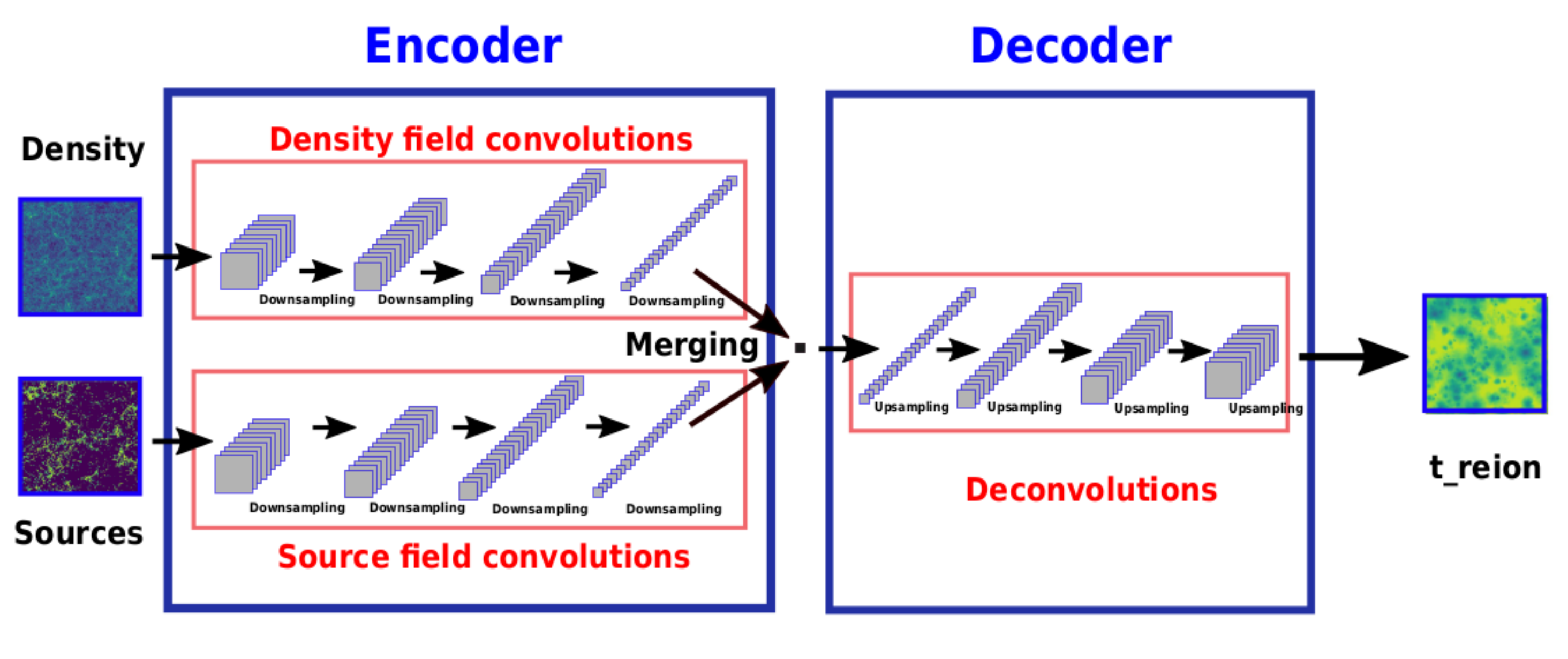}   
  \caption{Architecture of the convolutional auto-encoder model used to predict maps of $\mathrm{t_{reion}}$ (See Appendix \ref{appendix_1} for full details).
  The auto-encoder has two entries: the maps of the gas density field and of the star number density which are both Gaussian-smoothed (see Sect. \ref{dataset} for detailed explanations). There are two distinct blocks of convolution filters applied on both fields independently represented by the two red rectangles on the left. These two branches of convolution represent the encoder. After those independent series of convolutions, the outputs of the two last layers in the network are averaged together before entering the process of deconvolution (represented here as the third red rectangle which constitutes the decoder). At the end, one last layer with a linear activation function is applied to produce a full two-dimensional map of $\mathrm{t_{reion}}$ with continuous values.}
    \label{modelschematic}
  \end{center}
 \end{figure*}

In order to predict $\mathrm{t_{reion}}$ 3D maps, we use both the gas density (taken as the log of the baryon overdensity) and star particle number density, at z=6. This choice is arbitrary and driven by simplicity : gas density tracks the distribution of photons absorbers, whereas the star number density tracks the distribution of emitters. Note that the star number density is only an incomplete view of the photon production history : no information about the age or the emissivity is provided here. 
Moreover, we could also imagine different kinds of ionizing sources at play in a single simulation. This would entail Lyman continuum photon production phases that differ in length (see for example \citealt{2016MNRAS.456..485S} who have shown that binary interactions increase the length of the ionization photon bright phases of sources).
Evidently other choices would have been possible, possibly using more information, but as a proof of concept we will show that even this admittedly simple choice of inputs provides satisfying predictions at this stage. 

As explained hereafter, we will use a convolutional neural network  for our predictions, usually used for image processing in 2D. In theory, such networks can process 3D fields, such as an image with multiple channels, but then become quite memory consuming and less efficient, especially when the three dimensions are of commensurable sizes. 
Accounting for the limitations of the hardware currently available to us, we thus decided to make this first study using 2D CNN : gas and stellar number densities are provided to the CNN as 2D slices, and equivalently, predictions on $\mathrm{t_{reion}}$, are returned as 2D planes. Nevertheless, to capture some information along the direction normal to the plane, we Gaussian smooth the three-dimensional gas and stellar fields along this direction : we take a smoothing length of $\mathrm{\sigma=30}$ , corresponding to a size of 3.75 cMpc/h for the simulations studied here. We marginalized our search over different values for this smoothing scale, and the latter value gave us the best performance when training the neural network.
For a 3D reconstruction, all successive slices of a $\mathrm{t_{reion}}$ cube are predicted and stacked. Of course, this creates discontinuities along the stacking direction : to mitigate this effect, we perform three separate 3D predictions using this procedure, stacking along the three different main directions and combine them to obtain our final 3D prediction of $\mathrm{t_{reion}}$. Further details can be found in appendix \ref{appendix_2}.

From our $\mathrm{256^3}$ three-dimensional fields taken from TRAINSIM, we construct a sample of 3000 maps of $\mathrm{128\times 128}$ cells for the stellar and gas densities, and $\mathrm{t_{reion}}$.
This constitutes what we usually call the training set.
The maps are picked randomly inside the whole three-dimensional fields and the same location is taken for the three fields. 
In addition to the training set, we also build a test set composed of 500 additional $\mathrm{128\times 128}$ maps for the three fields, still from TRAINSIM. Such a test set is here to measure the accuracy of the trained model on unseen data during the training process. Therefore we ensure that the maps taken to build this test set are different from the ones belonging to the training set.   

Finally, we normalize the input in the neural network and we proceed as follows for both the stellar (S) and the gas density (D) fields:
\begin{itemize}
 \item{We take the mean of our fields in the whole training set: $\mathrm{<S>}$ and $\mathrm{<D>}$}
 \item{We subtract that value from all the values in the maps of the training set: $\mathrm{S=S-<S>}$ and $\mathrm{D=D-<D>}$}
 \item{We calculate the standard deviation of those new fields: $\mathrm{std(S)}$ and $\mathrm{std(D)}$}
 \item{We divide all the values of S and D by this value: $\mathrm{S=S/std(S)}$ and $\mathrm{D=D/std(D)}$}
\end{itemize}

Fig. \ref{ICsauto} shows an exemple of data used to train the neural network.
We show five different examples of both the transformed stellar and gas density fields that are the input of the network, and the corresponding $\mathrm{t_{reion}}$ field the network aims to predict.
We can see at first glance the correlation between the three different fields in each example case.
Therefore, we can expect that the network, with data that are transformed this way, should be able to infer the underlying correlation.

\subsection{Convolutional neural network architecture}
\label{architecturecnn}
 
Fig. \ref{modelschematic} shows a schematic view of the neural network used here (See Appendix \ref{appendix_1} for full details). 
The neural network we build is a special case of convolutional neural networks. It is called an auto-encoder and has the unique property of generating a complete image as an output.
An auto-encoder is a non recurrent neural network propagating forward (i.e. we process from  left to right in Fig. \ref{modelschematic}) with an input layer, an output layer and one or more hidden layers in-between.
An auto-encoder is always divided in two parts : the encoder and the decoder. 
The encoder is a succession of convolutions of the input of a layer with filters of a particular size. The results are then downsampled and given as inputs to the next layer. 
The decoder is the symmetric counterpart of the encoder.
Instead of convolution and downsampling, a layer is composed of a deconvolution plus an upsampling of the results. 

The auto-encoder we build here is special because it has more than one input image even if we aim to predict a single output image. In practice, we just apply the same series of convolutions and downsamplings independently for both our input stellar and gas density fields (the two distinct red rectangles on the left in Fig. \ref{modelschematic}). 
After the forward passing of these maps  through the layers of their dedicated branches of the encoder, the results are merged together (i.e. we take the average of the output maps of the gas density and the stellar branches) to feed the decoder. This input successively goes through the same number of layers as in the encoder. This eventually leads to a final map \dom{($\mathrm{t_{reion}}$)} of the same size as the starting inputs. 

We employ the usual Adam algorithm for the optimizer and we choose the mean squared error between the predicted and true two-dimensional maps  for the loss function to optimize. 
The choice of mean squared error is dictated by the regression nature of the problem we are facing here (i.e. prediciting continuous values of $\mathrm{t_{reion}}$ instead of discrete values) in contrast to classification problems.

\subsection{Training the neural network}
\label{training} 

\begin{figure}
\begin{center}
    \includegraphics[width=\columnwidth]{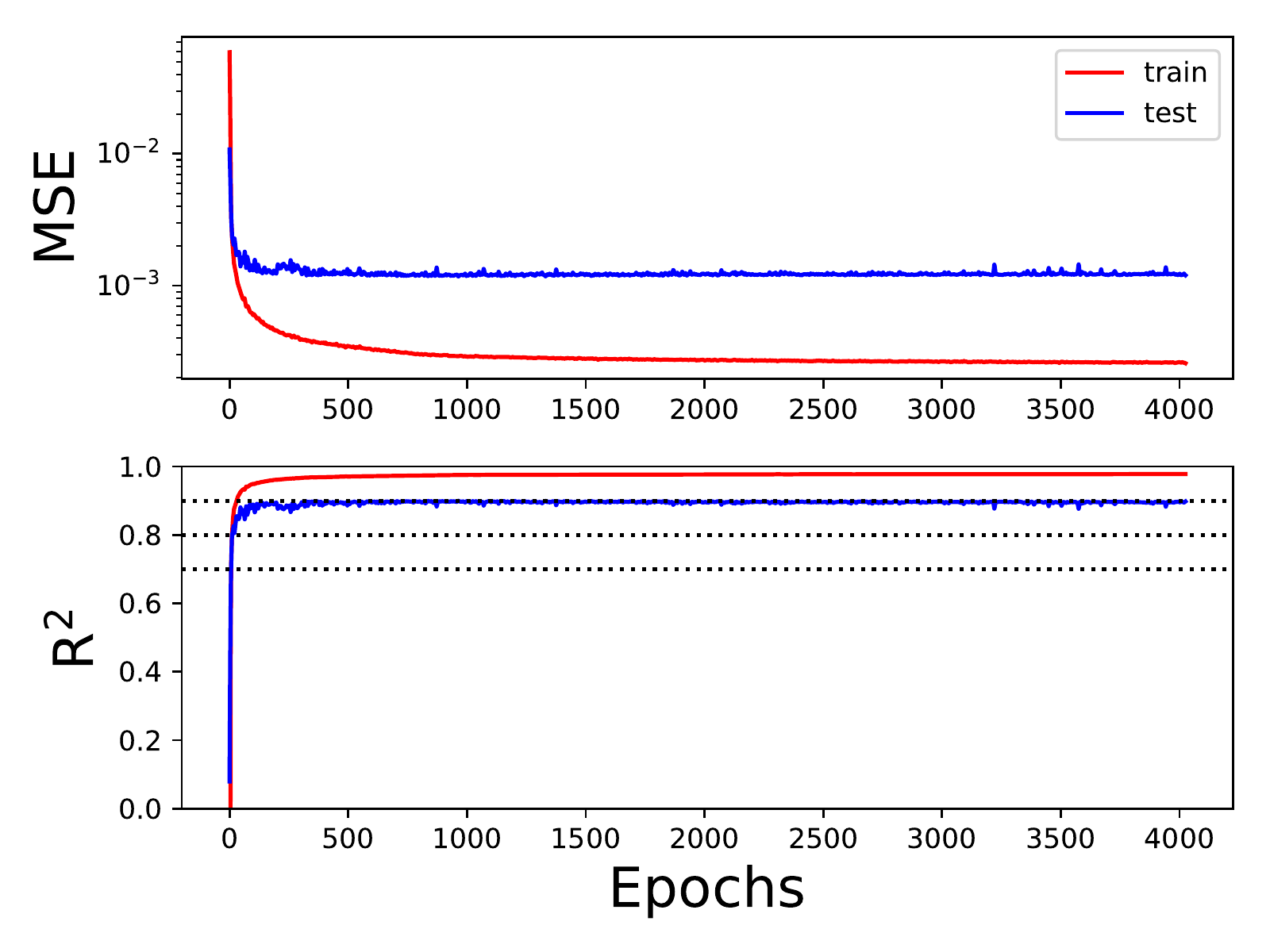}   
\caption{Training performance curves. The top panel shows the evolution of the mean squared error (MSE) between the predicted values and the real ones from the TRAINSIM simulation as a function of the number of epochs for both the training set and the testing set. This is the actual value of the loss function used to train the model and what the gradient descent algorithm is trying to minimize. The bottom panel shows the evolution with the number of epochs of the coefficient of determination $\mathrm{R^2}$ of equation \ref{r2equ}. It allows to monitor how our model matches the original values and in particular how it performs on unseen data with the testing set drawn from the TRAINSIM simulation. The different horizontal dashed lines show values of $\mathrm{R^2 = }$ 0.7,0.8 and 0.9.}
\label{curveloss}
\end{center}
\end{figure}

The auto-encoder described in the last section is built using the python API for neural networks KERAS\footnote{\href{https://github.com/keras-team/keras}{https://github.com/keras-team/keras}} (\citealt{2015xxxxcholkeras}).
Training the network is done on two Tesla K20 GPUs with the parallel training option of KERAS. 
Training on our setup takes about 28 hours with both GPUs when aiming to achieve best performance.
In principle, we can train our network on a much larger number of GPUs and therefore the wallclock time needed to train the network can be greatly reduced.

As already mentioned, to train a neural network, we build both a training set and a test set.
The training set, is made of data that are used to minimize the loss function.
On the other hand, the test set is made of data of the same nature as in the training set but are not used during the minimization process. They are only produced to control how a current version of the trained model performs on unseen data.
Therefore, we use both the training and the test set during training to monitor each training process.

To monitor our training performance we use two indicators. First we use the mean squared error (MSE) between the predicted $\mathrm{t_{reion}}$ and the true values. 
In practice, we want the MSE to decrease during the learning process until it reaches a plateau, indicating that the maximum learning potential has been achieved. 
However,  the MSE value is not meaningful taken in isolation, and does not tell us much about the quality of the predictions.   
The same MSE value can correspond to predictions of variable quality from one problem to another.

To measure the correctness of our prediction, we use a second indicator which is called the coefficient of determination $\mathrm{R^2}$ calculated with the following formula (see \citealt{2019MNRAS.484..282G}):
\begin{equation}
\mathrm{ R^2=\frac{\sum(y_{pred}-\overline{y}_{true})^2}{\sum(y_{true}-\overline{y}_{true})^2}  = 1 - \frac{\sum(y_{pred}-y_{true})^2}{\sum(y_{true}-\overline{y}_{true})^2}} 
\label{r2equ}
\end{equation}
In practice, a value close to 1 represents a 100 \% match between our original data and the ones predicted by the neural network.

The upper panel of Fig .\ref{curveloss} shows the evolution of the MSE as a function of the number of training epochs,
while the lower panel shows the evolution of $\mathrm{R^2}$. 
We show these curves for our best model, after we found the best way to pre-process our data, and the best network architecture with the best hyper-parameters. Trends are shown for the training set and test set, both from TRAINSIM. 
We clearly see the MSE decreasing quickly at the beginning of training for the first 250 epochs for both the training and the test set. This means that the choice of parameters is well-suited to the current problem and that the model learns efficiently. 
After about 250 epochs, the test set reaches a plateau while the training set keeps decreasing. 
We continue training up to the moment when the MSE curve reaches a plateau for the training set. The beginning of the training set plateau is generally considered as the moment when the best performances are achieved.
We achieve this after $\sim$ 2500 epochs.

Focusing on $\mathrm{R^2}$ in the bottom panel, we observe that the model reaches an accuracy of about $\mathrm{R^2\sim 0.99}$  on the training set when the MSE stabilizes.
Meanwhile the test set reaches a value of about $\mathrm{R^2 \sim 0.9}$ which means that our model generalizes well on unseen data.
However, both the training and the test set are built from the TRAINSIM simulation. Even if we make sure they are not the same maps, nothing guarantees that the model generalizes well on other, completely disconnected simulations.
That is why we ran the TESTSIM to test our model's performances on new data. All the results given in Sect. \ref{validation} will thus be given by applying the model to this TESTSIM simulation, unseen during the training.

\begin{figure*}
   \begin{center}
      \includegraphics[width=\textwidth]{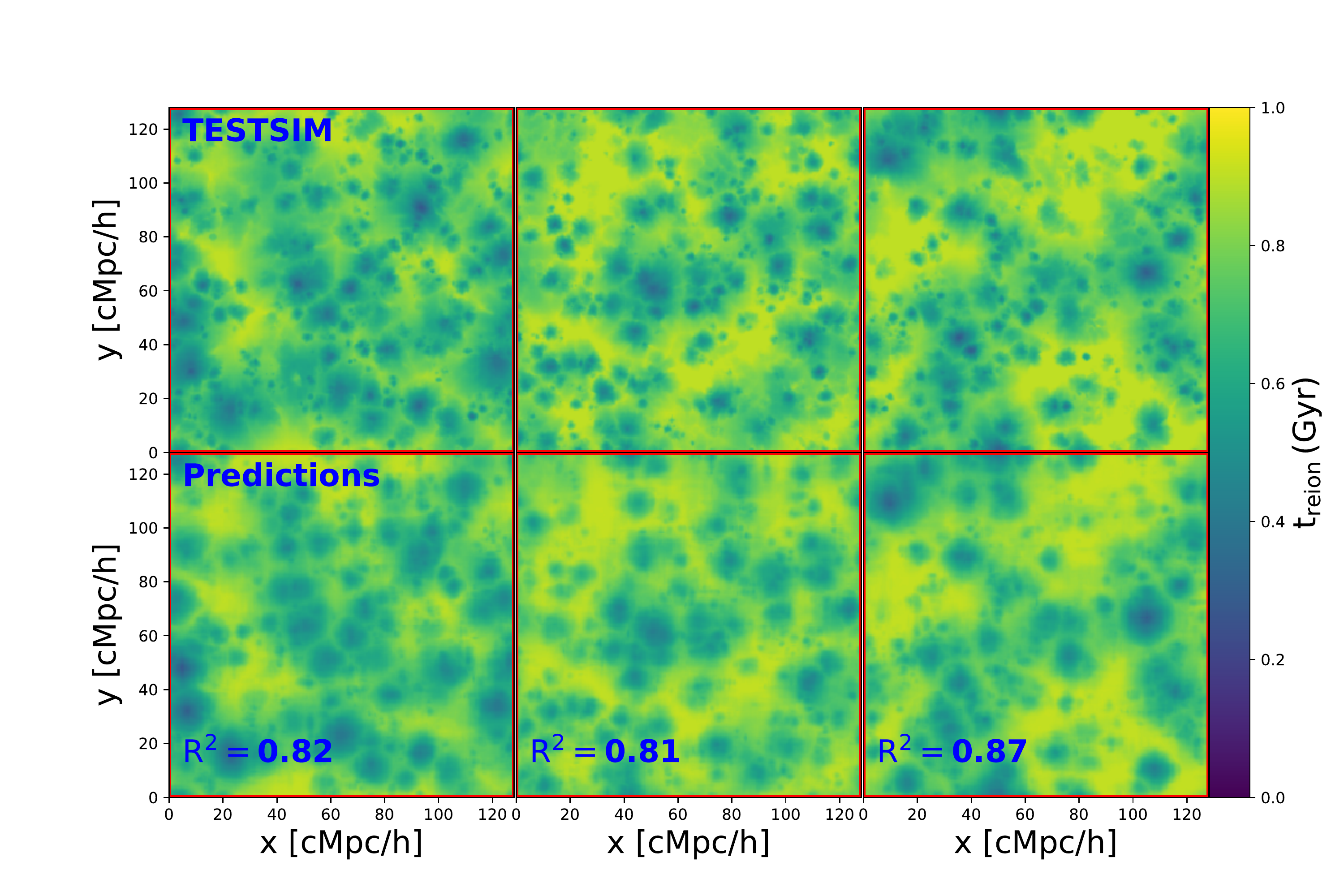}   
  \caption{Example of slices of $\mathrm{t_{reion}}$ from the TESTSIM simulation and the corresponding predictions from the model.
 The upper panel shows three different slices from the TESTSIM simulation, while the lower panel shows the prediction of the same slice with our best model. Average values of $\mathrm{R^2 \sim 0.83}$ are currently achieved with our best model. The predicted fields are reconstructed following the procedure described in Sect. \ref{cube_reconstruction}.}
    \label{field_reconstructed_three_slices}
  \end{center}
 \end{figure*}

\section{Results}
\label{validation}

\subsection{Field reconstruction}
\label{cube_reconstruction}

To measure the performance of our model on unseen data, we first use our trained network to reconstruct a field of $\mathrm{t_{reion}}$ in the TESTSIM simulation. We use the gas density and the stellar fields of this simulation and we transform them the same way we transformed the input training data (see Sect. \ref{dataset}).
We then predict mutiple two-dimensional slices to reconstruct the whole $\mathrm{t_{reion}}$ cube following the procedure described in Appendix \ref{appendix_2}.

Fig. \ref{field_reconstructed_three_slices} shows central slices of the reconstructed cube of $\mathrm{t_{reion}}$ in the three different directions. 
The model is well adapted for predicting $\mathrm{t_{reion}}$ maps in the three directions. 
The colormap in both the predictions and the original data is set to be the same to enable direct comparison.
Overall, the model predicts a range of continuous values of $\mathrm{t_{reion}}$ that are within the same range as the original simulation. 
Moreover, the global shape of the field seems to be well-predicted by the network which means that the large scale structures of the field seems to be well-learned by the model.

However, we report some differences at smaller scales. 
The network struggles to predict the exact same shape for the edges of $\mathrm{t_{reion}}$ bubbles.
Moreover, the peaks of small $\mathrm{t_{reion}}$ values are too high compared to the original data, meaning that the first sources episodes of reionization are only partially recovered.
Some small $\mathrm{t_{reion}}$ bubbles are missed during the reconstruction, or some spurious bubbles are created where they are not present in the original simulation.

All the aforementioned drawbacks of the model show its limited capacity to make robust predictions at small scales, which could be improved with a combination of a larger data set for the training and different kinds of inputs. For example, the star number density at z=6, is actually degenerate and can be similar for different source production histories : it is therefore not surprising that our model struggles to perfectly reproduce the time evolution. In fact, given this limitation, the ability of the network to predict a reionization timeline similar to the actual one can even be seen as surprising and indicates that to some extent,  the star production history is encoded in the gas density distribution and stellar number density. The inclusion of information about the source ages could surely improve the prediction on  reionizations' evolution, especially at early times.

Finally, in the lower left corner of each slice of the reconstructed fields, we also show the value of the $\mathrm{R^2}$ coefficient of prediction for the corresponding slices compared to the original slice of the TESTSIM simulation.
We reach average values of $\mathrm{R^2 \sim 0.83}$ over the three directions with our best model, without any fine tuning of the model.
Again, we expect to increase this accuracy with fine tuning of the hyper-parameters of the model, a bigger sample for the training, and better pre-processing of the input data.

\subsection{True versus pedicted values}
\label{True_vs_pred}

\begin{figure}
\begin{center}
    \includegraphics[width=\columnwidth]{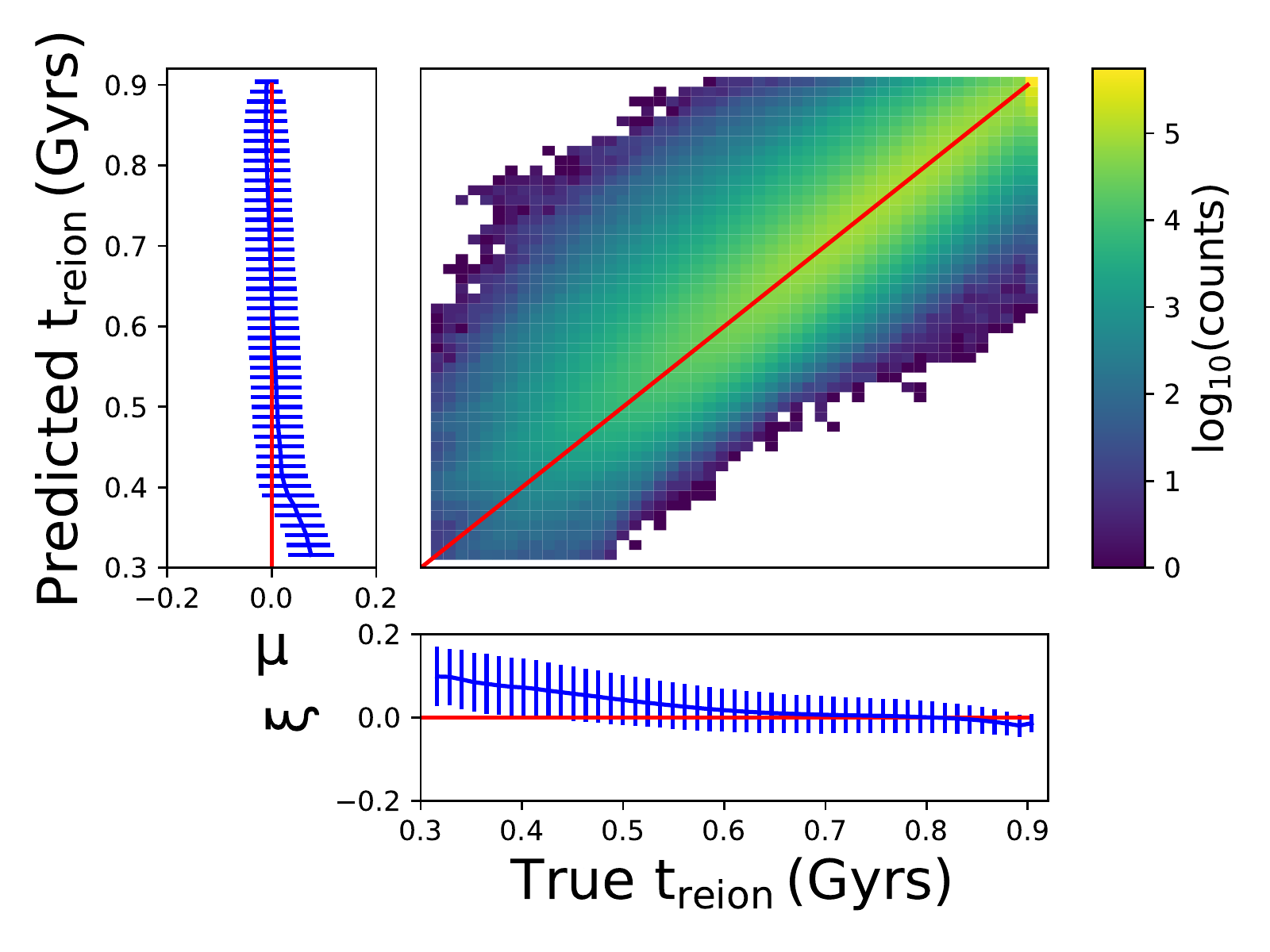}   
\caption{2D-histogram of the true versus predicted values of $\mathrm{t_{reion}}$ for our best neural network model once the training is finished. The histogram is constructed on the whole 3D reconstructed cube of the TESTSIM simulation as explained in Sect \ref{cube_reconstruction}. The red line shows the one-to-one relation while the color map encodes the number count of cells lying in the 2D space of true versus predicted. The bottom and left histograms show the mean and the standard deviation of the residual : $\mathrm{r =  Predicted \, - \, True }$ in the vertical and horizontal directions. The bottom histogram is the learning error, p(r\textbar True), while the side histogram is the recovery uncertainty, p(r\textbar Predicted).}
    \label{tr_vs_pred}
\end{center}
\end{figure}

As a second test, we also construct the 2D-histogram of the true versus predicted values of $\mathrm{t_{reion}}$. 
Fig. \ref{tr_vs_pred} shows the number count of cells lying in the true-predicted plane.
The red line on the figure shows the one-to-one relation. 
To highlight the differences, the bottom and left histograms show the mean and the standard deviation of the residual : $\mathrm{r =  Predicted \, - \, True}$ along the  vertical  and  horizontal  directions,  respectively. 

The left side histogram which displays the distribution of residuals as a function of the prediction (p(r\textbar Predicted)) is an actual measure of the uncertainties on predictions $\mu$.   
Such a test is standard in deep learning model validation, and the closer the distribution to the one-to-one relation, the better the performance of the network (see \citealt{2019MNRAS.484..282G}).  
Overall, our model tends to predict values of $\mathrm{t_{reion}}$ close to the one-to-one relation which once again demonstrates the ability of our network to perform this particular task.

We note that the average of the residuals in this case is well-centered on the zero residual value for $\mathrm{t_{reion} \ge 0.4}$ Gyrs. 
We report values of $\mathrm{\overline{\sigma} = 0.045}$ Gyrs for the mean of the uncertainty on prediction along all this range of $\mathrm{t_{reion}}$ values. 
This value is fairly constant over all this range and it means that we only have a 4.5\% error on our predictions on average. We measure a minimum value of  $\mathrm{\sigma = 0.010}$ Gyrs and a maximum uncertainty of $\mathrm{\sigma = 0.061}$ Gyrs indicating an error fluctuating between 1 and 6 \% compared to real values.
However, the model clearly overestimates the values of $\mathrm{t_{reion} < 0.4}$ Gyrs which means that we miss the first ionized regions in our predictions.

Finally, the bottom histogram, showing the distribution of residuals as a function of true values (p(r\textbar True)), represents the network error $\xi$.
The average value of the residual is well-centered around zero in the range [0.5-0.9] Gyrs for values of $\mathrm{t_{reion}}$.
We conclude that our network makes robust predictions in this range with a mean uncertainty of $\mathrm{\overline{\sigma} = 0.05}$ along these values. This is not surprising since these values of $\mathrm{t_{reion}}$ correspond to the large scales that are well-predicted in the maps of Fig. \ref{field_reconstructed_three_slices}.
However, we note that larger uncertainties are reported for $\mathrm{t_{reion}<0.5}$ Gyrs. In this case, the mean residual is significantly above the zero value, which suggests a larger predicted $\mathrm{t_{reion}}$ in this range. This can be seen in Fig. \ref{field_reconstructed_three_slices}, where these peaked locations in the maps have higher values in the predictions compared to the original data. Therefore, our model seems to struggle to predict the smaller scales in the maps, which correspond to the first locations in the simulation that were reionized by the first generation of ionizing sources.

  \begin{figure}
   \begin{center}
      \includegraphics[width=\columnwidth]{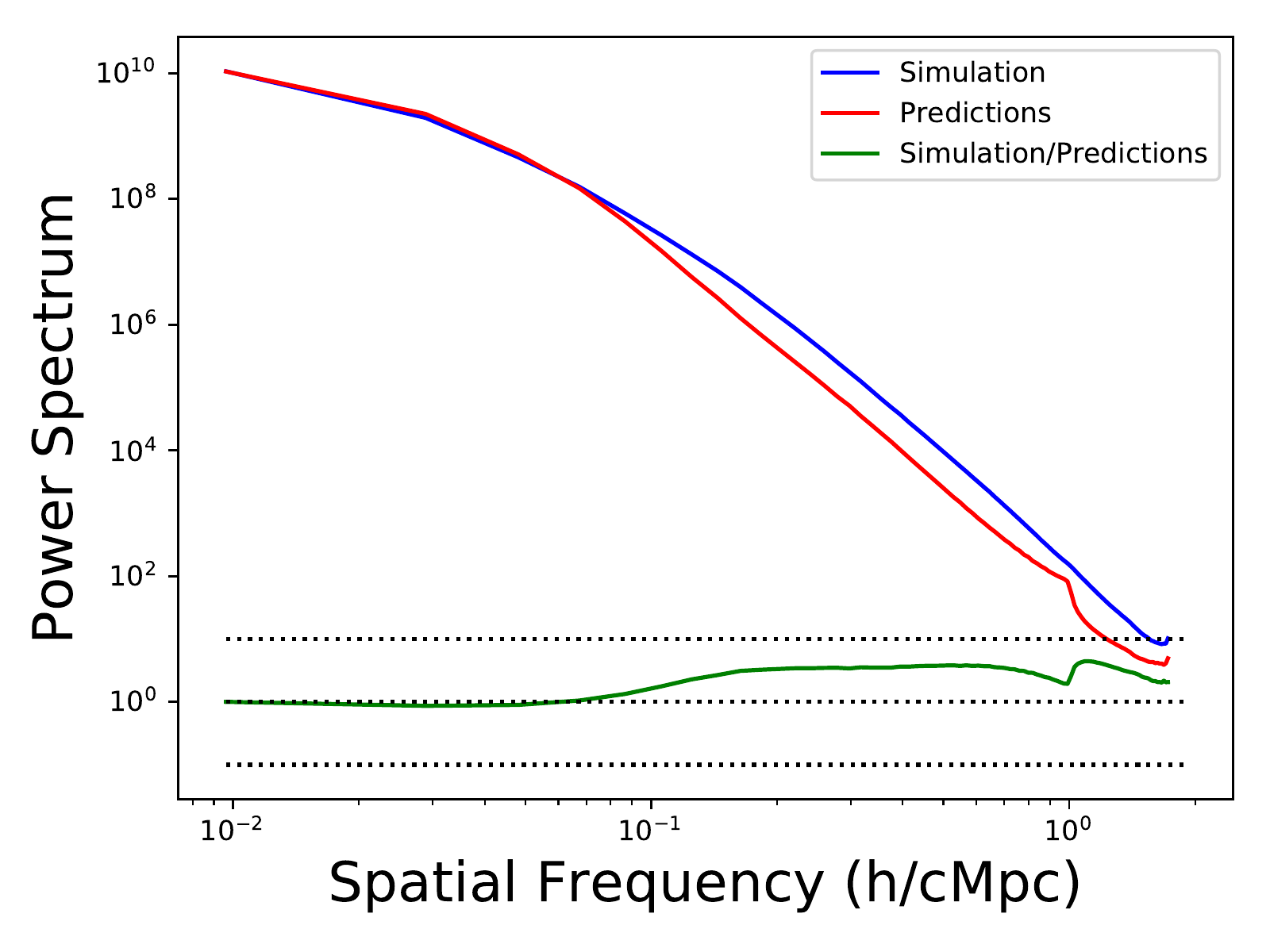}   
  \caption{Power spectra of the $\mathrm{t_{reion}}$ fields. The blue and red lines show respectively the power spectrum of the original TESTSIM simulation and the one predicted by our best model once the training is finished. Power spectra are computed on the whole three-dimensional cube. The green line shows the ratio of both power spectra.}
    \label{pspectrum_comp}
  \end{center}
 \end{figure}

\subsection{Power spectra}
\label{pspectra}

   \begin{figure*}
   \begin{center}
      \includegraphics[width=\textwidth]{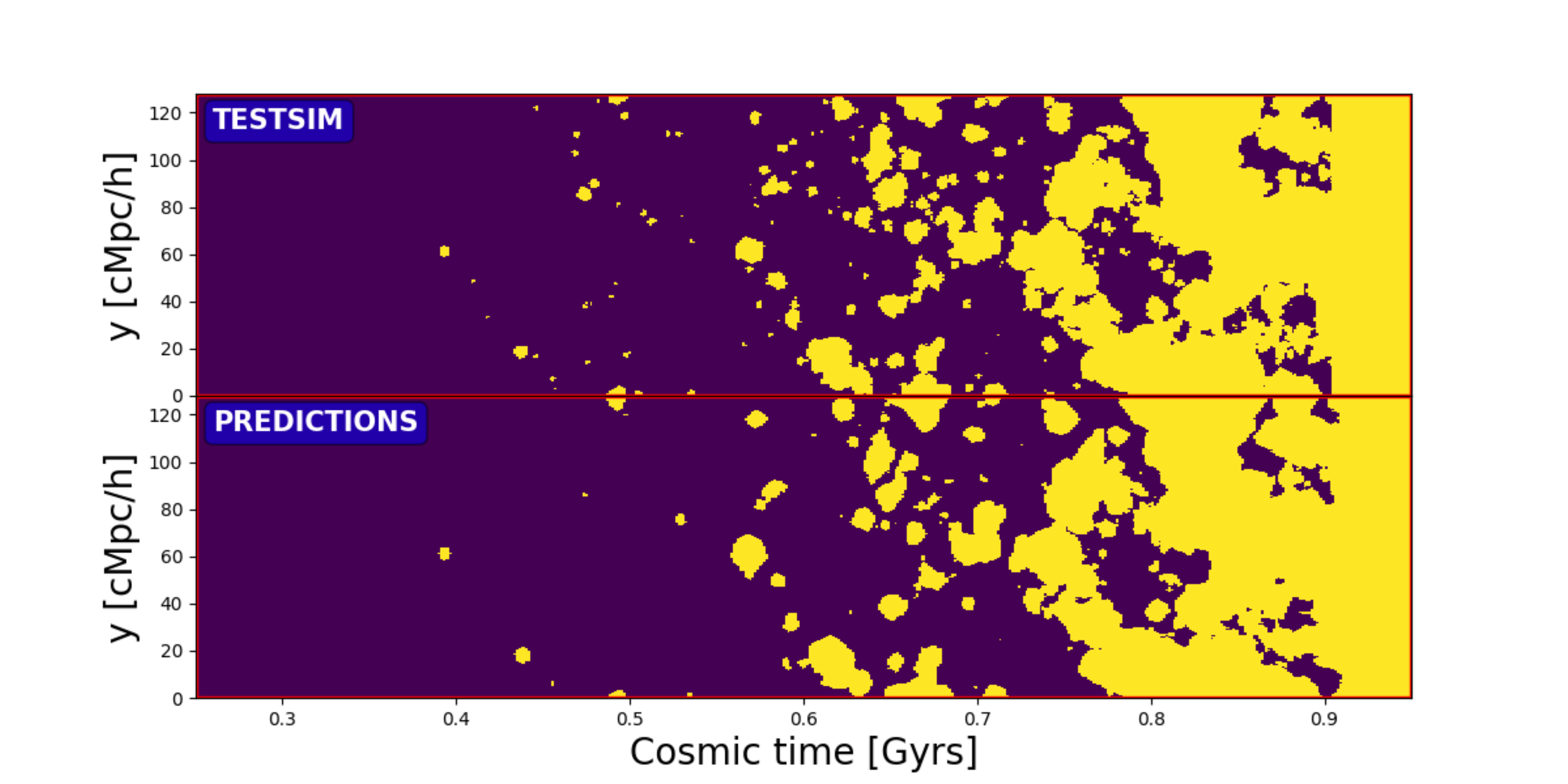}   
  \caption{Example of light cone from the TESTSIM simulation and the corresponding predictions from the model.
  The upper panel shows the lightcone of the central slice of the TESTSIM simulation, while the lower panel shows the prediction of the same slice with our best model.}
    \label{HII_dynamic_reconstructed}
  \end{center}
 \end{figure*}
 
As a third test, we also compute the 1D power spectrum $\mathrm{P_{1D}(k)}$ of the 3D field of $\mathrm{t_{reion}(\overrightarrow{r})}$ for both the original TESTSIM simulation and the reconstructed prediction of the network.
The power spectrum is defined as the azimutal average of the square of the module of the 3D Fourier transform $\mathrm{\delta(\overrightarrow{k})}$ of the 3D $\mathrm{t_{reion}(\overrightarrow{r})}$ field. This is computed as follows :
\begin{equation}
 \delta(\vec{k})=\int t_{reion}(\vec{r})e^{-2 \pi i\vec{k}.\vec{r}}d\vec{r}
\end{equation}
\begin{equation}
 P_{1D}(k)=<|\delta(\vec{k})|^2>_{|\vec{k}|=k}=\frac{\sum_{|\vec{k}|=k}|\delta(\vec{k})|^2}{\sum_{|\vec{k}|=k}1}
\end{equation}

Fig. \ref{pspectrum_comp} shows these two power spectra as well as the ratio of the two.

Overall, we report a perfect match between true and predicted values at large scales. The two spectra are on top of each other up to scale $\mathrm{k = 0.1}$ h/cMpc (i.e the ratio of both spectra is almost equal to one). This is in agreement with what is observed in Figs. \ref{field_reconstructed_three_slices} and \ref{tr_vs_pred} where large scale structure in the maps (with values of  $\mathrm{t_{reion}>0.5}$ Gyrs) are well-predicted by the network.

However, at scales $\mathrm{k > 0.1}$ h/cMpc the network seems to underpredict the power compared to the real simulation. Again, this discrepancy corresponds to small scales that are missed in the maps of Fig. \ref{field_reconstructed_three_slices} with $\mathrm{t_{reion}<0.5}$. This means that the network struggles to keep track of the first ionizing sources that appeared during the simulation.

\subsection{Reionization history}
\label{QHII_recovered}

 The previous sections show results on the prediction of the field of $\mathrm{t_{reion}}$, which encapsulates the whole reionization history of a given simulation. Here, we propose to use this map to demonstrate the potential of our model to predict the evolution of average quantities during the process of cosmic reionization, and how it compares with real simulations. We aim at showing how this could be useful for those who only have hydro simulations, and want to get an emulation of the radiative transfer calculation without performing it.   
 
 First, in Fig .\ref{HII_dynamic_reconstructed}, we show the reconstructed evolution with cosmic time of the ionized regions' expansion. To synthetize this evolution, we construct lightcones from $\mathrm{t_{reion}}$ slices.
 The upper panel shows the evolution of the lightcone constructed from the central $\mathrm{t_{reion}}$ slice of the TESTSIM simulation, while the lower panel shows the same evolution predicted by the model. Such fields are constructed as follows: 
 
 \begin{itemize}
 \item{We first take a slice of the 3D $\mathrm{t_{reion}}$ field.}
 \item{We consider a cosmic time of $\mathrm{t_{HII}}$ at which we want to create the HII regions' spatial distribution.}
 \item{We keep all the $\mathrm{t_{reion}<t_{HII}}$ cells.}
 \item{We mark them as ionized with a value of one.}
 \item{We consider the other cells as neutral with a value of zero.}
 \item{We repeat this for multiple values of $\mathrm{t_{HII}}$ and stack the results to construct the lightcone.}
 \item{We follow this procedure for both the $\mathrm{t_{reion}}$ field of the TESTSIM simulation and the prediction of the network.}
\end{itemize}

Overall the two lightcones look rather similar between the prediction and the TESTSIM simulation.
We observe better agreement at large cosmic times (i.e. when the reionization process is well-advanced).
The HII bubbles in the predicted field are at the right location with sizes comparable to the original ones.
However, the edges of the bubbles are somewhat different : some of them are merged in a single bubble in the predicted field, whereas several disconnected bubbles are reported in the original data.  We also observe some bubbles completely disappearing in the prediction compared to the TESTSIM simulation. The inaccuracy of the model at predicting the smallest scales is reflected here when predicting the HII regions' spatial distribution as a function of cosmic time.
 
At early times, for $\mathrm{t\sim0.5}$ Gyrs, some of the first and smallest bubbles are missing.
This illustrates again the fact that the model struggles at predicting the first stages of the HII regions' expansion.
This is due to the fact that the model is unable to predict the smallest $\mathrm{t_{reion}}$ values in Fig. \ref{field_reconstructed_three_slices} which reveals the limitations of the model to get accurate apparition times for the first generations of ionization sources early on in the simulation.

\begin{figure}
   \begin{center}
      \includegraphics[width=\columnwidth]{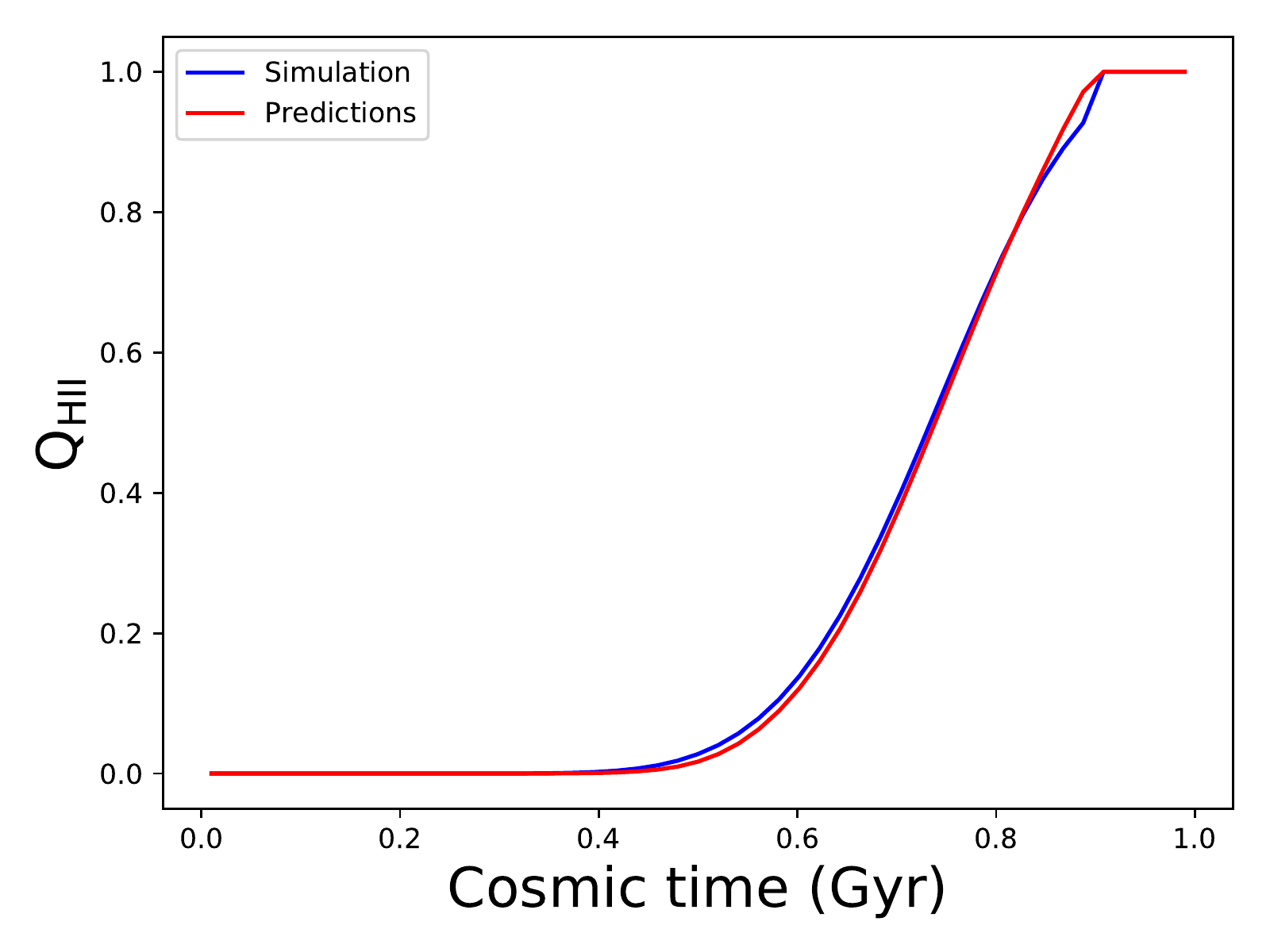}   
  \caption{Evolution of the volume filling factor of HII regions $\mathrm{Q_{HII}}$ with cosmic times in both the original simulation (TESTSIM) and the prediction of the neural network. $\mathrm{Q_{HII}}$ is calculated from the whole 3D field of $\mathrm{t_{reion}}$ in both cases by taking the cumulative sum of the histogram of the $\mathrm{t_{reion}}$ values.}
    \label{QHII_comparison}
  \end{center}
 \end{figure}

As a second test, we compute the evolution of the fraction of the volume that is ionized at a given cosmic time. In Fig .\ref{QHII_comparison}, we present the evolution of the volume filling factor of HII regions $\mathrm{Q_{HII}}$ in both the TESTSIM simulation and the model prediction. The evolution of this quantity is calculated by binning the cosmic time period and by getting the cumulative sum of ionized cells at a given cosmic time from the $\mathrm{t_{reion}}$ field. We report an almost perfect match of the two curves in Fig .\ref{QHII_comparison} at all cosmic times.
This demonstrates the ability of our model to predict a global reionization history during the whole simulation.
We observe some minor differences at cosmic time $\mathrm{0.4\le t\le0.6}$, where $\mathrm{Q_{HII}}$ of the TESTSIM is somewhat above the model prediction. Once again, this is due to the inability of the model to perform at predicting the smallest scale in the $\mathrm{t_{reion}}$ field.
Overall, our deep learning model is already well-designed to emulate a global reionization history, which can 
be useful for a wide range of studies.

 \section{Discussion and conclusion}
 \label{discussion}
 
 In this section we present and discuss our global results with their successes and drawbacks. 
 We finally conclude with the implications of such a study for the near future and what it may imply for the future of numerical simulations.

 \subsection{Successes}
 \label{success}

With this study we have demonstrated that deep learning models can emulate the physics of the radiative transfer occurring during the reionization epoch of the Universe. We used an auto-encoder neural network, usually designed for data compression, to create a generative model that predicts the whole reionization history encapsulated in a single field : the map of reionization times $\mathrm{t_{reion}}$. The built network takes the stellar number density as well as the gas density field from a simulation at the end of the reionization as inputs and produces the $\mathrm{t_{reion}}$ field as an output. 

With our current optimization strategy, we achieve a determination coefficient $\mathrm{R^2\sim 0.83}$ in recovering the whole three-dimensional field of $\mathrm{t_{reion}}$ on a test simulation that was never seen during the training of the network. The model was therefore successful at reproducing a variety of scales in this field down to $\mathrm{k= 0.1}$ h/cMpc which is already useful for a wide range of studies.
Moreover, the model has shown its ability to generate HII bubble dynamics in good agreement with real simulations with full radiative transfer. It gives at the end a mean reionization history almost identical compared to real data.

For the time being, we have therefore proved that an auto-encoder neural network architecture for emulating radiative transfer during reionization is a promising solution.
This is really encouraging since the prediction of a complete $\mathrm{256^3}$ $\mathrm{t_{reion}}$ cube is produced in a much faster way than a complete reionization simulation that includes radiative hydrodynamics.
The prediction of the total $\mathrm{256^3}$ cube is done in about 3 minutes on a single CPU.
The prediction could be even faster than this if we parallelize the reconstruction process of the whole cube, and optimize the current python API. 
In comparison, post-processing radiative transfer in the same way as in \citet{2015MNRAS.453.2943C} for a simulation with the same resolution and the same number of cells would be done in about one hour on a single last generation Pascal P100 GPU.

 \subsection{Caveats and road to improvements}
 \label{caveat}
 
The training of the current neural network is not perfect, but we expect a great scope of improvement.
First, we have shown that the predicted $\mathrm{t_{reion}}$ field is inaccurate at small scales.
This is related to the inability of the model to predict the smallest values of $\mathrm{t_{reion}}$ which correspond to the 
location where the first ionizing sources appeared during the TESTSIM simulation. 
This is not surprising as the current model is trained with the gas density and the stellar number density fields taken at the end of the simulation, at z=6.
Therefore, these data used as inputs of the model do not hold any temporal information about the ionizing source history through the whole simulation. However, the $\mathrm{t_{reion}}$ field we aim to predict is a summary of this integrated history. We could therefore expect to gain much accuracy in the prediction if we introduce such temporal information when training the network. For example, we could imagine taking a third field as an entry of the network, which could be the average cosmic time of apparition of the ionizing sources inside each cell troughout the whole simulation. In practice, adding such a third field shouldn't prove troublesome when getting access to a larger number of GPUs for the training phase of the neural network.

We also expect improvement by increasing the quantity of data used to train the network.
Indeed, the current performance was achieved with training on only a sample of 3000 independent images for the training set. It is well known that increasing the training sample increases the performance and usually training sample of the order of 80000 images (see \citealt{2019MNRAS.484..282G} for example) are used, which is much higher than what we currently have.
Moreover, the training set was built from a single simulation.
However, building the training sample from different simulations instead of a single one should improve the performance. Indeed, the current model can somewhat overfit in learning a biased representation of the density field for the particular initial conditions used for the simulation taken for the training.

We also expect some improvement with the properties of the network itself. 
First of all, the optimization of hyper-parameters was only briefly investigated in the current training of the model.
We can expect to get even better performance by focusing more on hyper-parameter tuning.
Other choices for the loss function to minimize could also be investigated such as a customized loss function tuned to perform this particular task.
Moreover, we could also reconsider our network architecture. We could imagine adding layers and changing the number of filters in each layer. The size of the convolution filters in each layer could also be changed, and systematic studies for tuning these parameters could be investigated more carefully.

Furthermore, the predictions of our neural network are currently done in two dimensions, as we are limited by hardware considerations.
Therefore we had to smooth the gas density field and the stellar field in the transverse direction to the plane we are trying to predict. However, we could expect better performance of the network by using three-dimensional convolutions instead of two-dimensional as currently done, providing a direct prediction for three-dimensional cubes and getting rid of the additional process of reconstructing the volume from two-dimensional maps.

 Finally, for the time being, the neural network has only been trained on a given simulation, with a particular box size and resolution as well as with a specific set of parameters for the input physics. Therefore, the network trained this way is only adequate when predicting $\mathrm{t_{reion}}$ maps on top of hydro-simulations with the same properties. We briefly tested the prediction of $\mathrm{t_{reion}}$ maps associated with simulations with different parameters with the current version of the network. Unsurprisingly, the model failed to perform satisfyingly on simulations with different parameters compared to the one  used for training.
 However, We see clear avenues of improvement to pursue, and hope to train networks that could perform on a variety of hydro-simulations with a large range of parameters.
 For the time being, we have shown that the methodology is feasible, and we plan to work on more generic networks for forthcoming studies. 

\subsection{Conclusion}
\label{conclusion}

With the present  study, we lay the groundwork for developing emulators of reionization simulations.
We have shown that deep learning methods should help to emulate realistic simulations very efficiently. 
Such techniques could considerably speed up our way to predict the ionization field associated with a hydrodynamic simulation, when compared to full radiative transfer calculations. The model presented in this study still suffers from disparities with actual simulations, but we expect great possibilities of improvement.
Of course, training such models requires a large sample of existing simulations, but many of these simulations have already been run by the community and constitute a data base that could be used for the systematic training of neural networks (see \citealt{2011MNRAS.414.3458W}, \citealt{2012A&A...548A...9C}, \citealt{2014ApJ...793...29G}, \citealt{2014ApJ...793...30G}, \citealt{2015MNRAS.453.2943C}, \citealt{2016MNRAS.463.1462O}, \citealt{2018arXiv181111192O} and \citealt{2018MNRAS.479..994R}, among others). 
We plan to use large sets of existing simulations with different parameters and different sizes and resolutions, thus aiming at creating a data base of networks that could be used by the community to emulate the radiative transfer on a variety of hydro-simulations with different parameters. 
Finally, a long term objective would be to end up with a neural network model that could emulate a complete simulation at once using only the initial conditions of the original simulation. Such an idealistic network could perhaps emulate all the ingredients of cosmic reionization simulation at once : the dynamic of dark matter (see e.g. \citealt{2018ComAC...5....4R}), the hydrodynamic of the gas (see e.g. \citealt{2019arXiv190412846Z}) and the radiative transfer.  More realistically, one could imagine replacing specific modules within existing simulation codes with trained neural networks.
With the database of simulations currently at our disposal, and the promise of more to come, a very exciting time for deep learning science applied to cosmology is upon us.

\section*{Acknowledgments}

We thank contributors to SciPy, Matplotlib, pyDOE, and the Python programming language.
We thank the KERAS and Talos API for deep learning machinery and optimization in Python. This work was granted access to the HPC resources of CINES under the allocation 2019- A005041061 attributed by GENCI.

\bibliographystyle{mnras}
\bibliography{biblio}

\appendix

\begin{table*}
  \centering
   
  \begin{tabular}{|c||c|c|c|c|}
  \hline
   Network branch & Layer \#/step name & Number of filters/data & Filter size/data dimension & Activation function \\
  \hline
  \hline
    & Input & 3000 & $128\times 128$ & . \\
    & 1 & 32 & $3\times 3$ & Relu \\
  Encoder    & 2 & 64 & $3\times 3$ & Relu \\
    & 3 & 128 & $3\times 3$ & Relu \\
    & 4 & 128 & $3\times 3$ & Relu \\
  \hline
  \hline
    & 1 & 128 & $3\times 3$ & Relu \\
    & 2 & 128 & $3\times 3$ & Relu \\
   Decoder & 3 & 64 & $3\times 3$ & Relu \\
    & 4 & 32 & $3\times 3$ & Linear \\
    & Output & 3000 & $128\times 128$ & . \\
  \hline
\end{tabular}

  \caption{Details of the architecture the auto-encoder convolutional neural network used to predict maps of the cosmic reionization time $\mathrm{t_{reion}}$. Each row shows the properties of a given layer in the encoder or the decoder. The different columns show different properties of the corresponding layers. The layers of the encoder are applied both to the input composed of the gas density and to the stellar field (See Fig. \ref{modelschematic} and Sect.\ref{architecturecnn} for explanations). The outputs of the encoder are averaged before entering the layers composing the decoder.}
  \label{network_architect}
\end{table*}

\section{Autoencoder architecture details}
\label{appendix_1}

Here we detail the architecture of our auto-encoder neural network.
Our neural network is composed of a series of four hidden layers for both the encoder and the decoder.
Table \ref{network_architect} gives the details on the number of filters and their size in all four hidden layers. 
To improve our model architecture, we also add what we call skip connections in the decoder part. Skip connections are here to add information when inputting maps into a layer of the network. In practice, it is a merging of multiple layer outputs to construct an input which is not only the output of the previous layer. In our case, we merge the outputs of every layer in the decoder with the corresponding outputs in the encoder in both the gas density and the stellar branches. It allows us to combine information from the current decoded version and the one at the corresponding step in the encoded version. In practice, training with skip connections improves the results compared to training without (\citealt{DBLP:conf/nips/MaoSY16}).

To avoid overfitting during the training (i.e. the fact of achieving a good fit for our model on the training data, while it does not generalize well on new, unseen data), we also add batch normalization plus dropout regularization right after the convolution/deconvolution respectively in the encoder/decoder in each layer. Batch normalization is a  transformation that maintains the mean output of a layer close to zero and its standard deviation close to one.
Dropout regularization is performed right after batch normalization and is what prevents overfitting.
Dropout regularization is activated through a value between zero and one that corresponds to a probability to shut down certain neurons (i.e. a given filter in a given layer). The fact of randomly shutting down neurons at every epoch is known to improve the accuracy of the model on unseen data (\citealt{2019arXiv190413310}).

We use the Talos\footnote{\href{https://github.com/autonomio/talos}{https://github.com/autonomio/talos}} tool with KERAS to tune our hyper-parameters. Talos allows to marginalize over the hyper-parameter space and gives correlations between them. In our case, we only marginalize over the learning rate and the dropout regularization values. We delay further improvement on hyper-parameters to upcoming studies as we want to highlight the proof of concept of predicting $\mathrm{t_{reion}}$ maps in the current study.

Finally, we use the usual Relu activation function in every layer except in the last one.
We use the Linear activation function in the last layer because we want to predict continuous values as an output of the network instead of discrete ones. This is different from common neural networks that aim to predict discrete values for classification problems.

\section{$\mathrm{t_{reion}}$ cube reconstruction}
\label{appendix_2}

Here, we describe our procedure to reconstruct a complete three-dimensional $\mathrm{t_{reion}}$ cube with our network.
Our neural network predicts two-dimensional slices of size 128 $\times$ 128 cells from maps of the gas density and the stellar density fields with the same size.
However, we choose to only keep the central submap of size 64 $\times$ 64 cells from a complete 128 $\times$ 128 prediction. This procedure ensures we do not miss sources nearby that are just outside the slice we are trying to predict. 
Therefore, we need to make 16 predictions to reconstruct the whole 256 $\times$ 256 slice for the simulations studied here.

To reconstruct the whole  $\mathrm{256^3}$ cube, we repeat the two-dimensional reconstruction of a slice 256 times.
Since the stellar and density fields used as inputs of the neural network are smoothed along the transverse direction to the plane we aim to predict, we reconstruct the cube by piling up reconstructed slices along this particular direction. 
However, in practice, it generates spurious grid artefacts when looking at slices taken along directions different from this direction. Therefore, instead of reconstructing the cube along only one particular direction, we reconstruct three different cubes along the three main directions. Then, we take, for each cell, the minimum and maximum values of $\mathrm{t_{reion}}$ from these three cubes. We then take the average of these two values to get our final three-dimensional reconstruction. Such a procedure has the advantage of eliminating the grid artifacts during the reconstruction procedure.

\end{document}